\documentclass[11pt,dvips]{article}
\usepackage{epsfig,times} 
\usepackage{wrapfig}
\makeatletter
\renewcommand{\section}{\@startsection{section}{1}{0in}
	{0.4\baselineskip}{0.1\baselineskip}{\Large\bf}}
\renewcommand{\subsection}{\@startsection{subsection}{2}{0in}
	{0.25\baselineskip}{-\baselineskip}{\large\bf}}
\renewcommand{\subsubsection}{\@startsection{subsubsection}{3}{0in}
	{0.1\baselineskip}{-\baselineskip}{\normalsize\bf}}
\makeatother
\begin{document}
%
\thispagestyle{myheadings}
%
\markright{HE 2.5.08}
\begin{center}
%
{\LARGE \bf Gamma Hadron Discrimination with a Neural Network 
\\ 
\vspace{0.3 cm}
            in the ARGO-YBJ Experiment}
\end{center}

\begin{center}
%
%
{\bf Presented by S. Bussino$^{1}$ for the ARGO-YBJ Collaboration}\\
{\it $^{1}$Dipartimento di Fisica and INFN, Universit\'a di Roma Tre, 
I-00146 Roma, Italy}
\end{center}

\begin{center}
{\large \bf Abstract\\}
\end{center}
\vspace{-0.5ex}
%
%
The structure of a neural network developed for the 
gamma hadron separation in the ARGO-YBJ detector is presented.
The discrimination power in the full ARGO-YBJ energy range is shown 
in detail and the improvement in the detector sensitivity is also 
discussed.
%

\vspace{1ex}

%
%
%
%

\section{Introduction}
\label{intro.sec}

The ARGO-YBJ detector (Abbrescia et al., 1996), described
elsewhere at this conference (D'Ettorre et al., 1999), is a full coverage 
$71 \ \times \ 74 \ m^{2}$ 
RPC carpet that will be installed at YanBaJing Cosmic Ray Laboratory
(Tibet, China) at an altitude
of $4300 \ m$ a.s.l. Due to the high altitude of the site 
and to the high granularity of the carpet, made of $14040$ pads 
$56 \ \times \ 62 \ cm^{2}$ each (dead space $\simeq \ 5 \ \%$), 
the electronic shower image can be used to discriminate between gamma and
proton initiated showers. 
In order to develop a neural alghoritm able to perform such a discrimination
more then $7 \cdot 10^{5}$ showers, 
generated using the CORSIKA package, were propagated through the 
detector and the neural net was structured around the 
peculiarities of each class of showers. 
The neural net performances have been studied for different sets of simulated
data and the results lead to a relevant increase in the sensitivity
of the detector.

\section{The Monte Carlo data sample}
\label{data.sec}

The events have been generated by using CORSIKA code 5.61 (Heck, 1998)
which provides a complete simulation of the shower development in the 
earth's atmosphere. The electromagnetic part of the shower simulation is 
realized by the EGS4 code (Nelson, Hirayama and Rogers, 1985), while for 
the adronic component the data have been processed by
VENUS code (Werner, 1993) for the high energy
hadronic interactions and by GHEISHA code (Fesefeldt, 1985) for the low
energy hadronic interactions (Knapp, Heck and Schatz, 1996). 
The data have been generated in 
the energy range $100 ~ GeV  - ~ 10 ~ TeV $ with an energy 
distribution given by:
\begin{equation}
N(E)dE \ = N_{0} \ E^{-2.7}dE  
\label{eq:powerlaw}
\end{equation}
Data have been generated in different energy intervals,
as shown in table
\ref{tab:cor2}. 
The Monte Carlo data have been processed through a code simulating 
the ARGO-YBJ detector, taking into account the energy loss
in the material, the ionization process in the RPC gas
and the RPC efficiency  ($94 \ \% $)
as measured during the test at YBJ (Mari et al., 1999). Also the background has 
been simulated and a basic trigger logic has been implemented, requiring at 
least $25$ signals inside the trigger time window.
For vertical showers simulated with the core at the center of the apparatus,
the number of events passing the trigger logic are reported in 
table \ref{tab:cor2}. 
\begin{table}[h]
\centering
\begin{tabular} {c c c c c } 
\hline \hline 
Mean Primary C.R. Energy & 
\multicolumn{2} {c}  {Number of generated showers} & 
\multicolumn{2} {c}  {Number of detected events}   
\\ \ \ $(Tev)$ \ & \ \ \it{gamma} \ \ & \ \ \it{proton} 
& \ \ \it{gamma} \ \ & \ \ \it{proton} 
\ \  \\ \hline 
$0.100<E<0.500$ & 360000 & 360000 & 37000 &   6000 \\
$0.500<E<1$ & 12000 & 12000 & 11000 &  10200 \\  
$1<E<3$ & 5000& 5000 &  4700 &   4200 \\  
$3<E<10$ & 770& 770  &   750 &    735 \\   
\hline \hline 
\end{tabular}
\caption[]{Number of generated showers in different window
energy of the primary particle.}  
\label{tab:cor2}  
\end{table}

\begin{figure}[t]
\centerline{\epsfig{figure=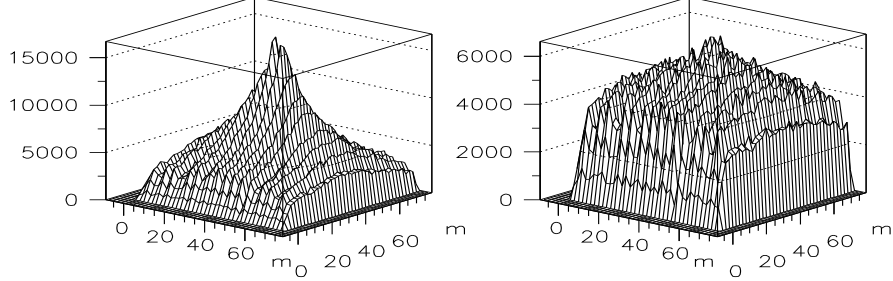,width=13.cm,height=4cm}}
\caption[]{The hits distribution in the  detector, for
vertical events with the core at the centre of the carpet, for
gamma (left) and protons (rights). Data are averaged over all the generated 
events.} 
\label{fig:nhitsxy}
\end{figure}

In order to plan the neural network architecture, we must know the general
behavior of gamma and hadron events, as a function of the observables
detected in the apparatus. The two classes of showers
can be distinguished by the different radial profile and from the
presence, in the hadronic showers, of local fluctuations induced by small
electromagnetic subshowers produced by neutral pion decay.
The distribution of the hits number as a function of the coordinates is shown 
in fig. \ref{fig:nhitsxy}: the two classes of events, averaged over the whole
energy range and over all the generated events, are clearly different. The 
fluctuations too are used as a tool in distinguishing
gamma- from hadron-induced showers by evaluating the 
quantity $F(x,y)$ defined as:

\begin{equation}
 F(x,y) \ = \frac{ | \ (N_{hits} - <N_{hits}> | \ }{\sqrt{<N_{hits}>}}  
\label{eq:fluct}
\end{equation}

\section{The neural network setup}
\label{network.sec}

The distribution shown in the previous picture (see fig. 
\ref{fig:nhitsxy}) are averaged over many events: the purpose of the 
neural network is to distinguish between the two classes event by event, 
and not only when all the fluctuations are smoothed by the statistics.
For this reason the network was planned looking for functions of observables 
which can discriminate the two classes and then the net 
architecture was implemented in such a way to combine the discrimination 
power of these functions to obtain the final rejection factor. 
We implemented a very simple neural network, taking into account that the
number of free parameters is strongly correlated to the number of Monte Carlo 
events needed to teach the network. 

\begin{wrapfigure}{r}{5.5cm}
\centerline{\epsfig{figure=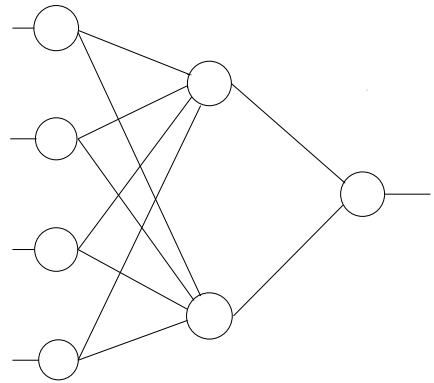,width=5.cm,height=5.cm}}
\caption[]{The setup of the network.}
\label{fig:figrete}
\end{wrapfigure}

The neural network (see fig. \ref{fig:figrete})
is a standard three layer perceptron. The input layer consists of four 
neurons, each reading four different functions of the observables: the intensity of the signal in 
each pad and its square, the quantity $F$ described by eq. \ref{eq:fluct} and 
its local maxima. Each neuron has a number of inputs corresponding to the pad
number in the detector, plus a bias signal to set the threshold. We divided 
the electronic image of each event in $1 \ m$ wide circular coronas, centered
around the core of the shower and symmetrized the neural weights within each
corona: all the inputs connected to the pads belonging to the same corona 
have the same weight. In this way, based upon the intrinsic symmetry of the 
whole physical process, the weight number is reduced to a maximum of
$145$ (corresponding to a shower with the core centered in one corner of the
detector) instead of $\simeq \ 14000$ (one for each pixel). The aim of the hidden and of the output layer is to 
improve the rejection factor by combining the outputs of the 
input layer neurons. We implemented a two neurons hidden layer, 
which seems a good compromise between flexibility and power. The output 
neuron furnishes the final result and the classification of the shower 
(i.e. $0$ for gamma induced showers and and $1$ for proton induced showers). 
\vspace{0.3cm}
\section{Results on gamma hadron separation}
\label{results.sec}

\begin{figure}[b]
\centerline{\epsfig{figure=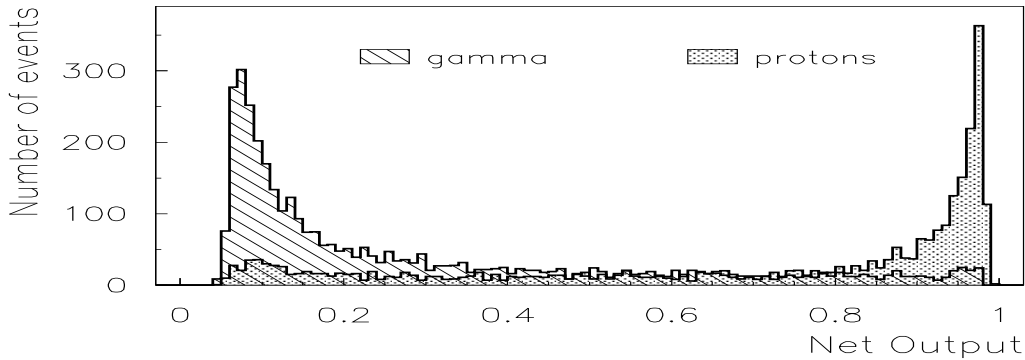,width=15.cm,height=4.8cm}}
\caption[]{The answer of the neural network is shown for events 
with $100 \ < \ n_{pad} \ < \ 150$.} 
\label{fig:reteout}
\end{figure}

The Monte Carlo events have been classified in several groups according to
intervals in the number of hits recorded in the apparatus, which is an 
observable. For each class of events of a different hit number, we could 
then tune the neural network in order to maximize its 
discrimination power. The data 
sample was processed using half of the available data to teach the net and 
the other half to evaluate the rejection power of the net. The descrmination
power of the network was initially evaluated for vertical showers with
the core at the centre of the detector. The results are summarized in the 
third and fourth colummns of table \ref{tab:results}, where it is shown
the fraction of particle rightly identified. We tested the effect of the
uncertainities on the core reconstruction, using very simples alghorithms for
the core position determination, and we found that the discrimination
factors $\epsilon$ are of the same order of those obtained without taking
into account the error reconstruction. 
This result can be easily understood looking at 
the radial profile of the weights of the input layer neurons. The slope
is smooth and the hit density 
is symmetric with respect to the shower core. Because of that, a small shift
due to the core position reconstruction has a small influence in the network
result. An example of the output of the neural network is given in
figure \ref{fig:reteout}, for events with a number of pads in the
range $100 \div 150$.
\begin{table}[t]
\centering
\begin{tabular} {c c c c c c } 
\hline \hline 
Hits Number & Total Number & 
\multicolumn{2} {c} {Central Core Events} &
\multicolumn{2} {c} {Edge Core Events} 
\\  &  \ of Events \ & \ \ $\epsilon_{\gamma}$ \ \ & \ \ $\epsilon_{h}$ 
\ & \ \ $\epsilon_{\gamma}$ \ \ & \ \ $\epsilon_{h}$
\ \  \\ \hline 
$50  \le n \le 75$    &  4000   &  75 \% & 74 \% &  75 \%  & 75 \%  \\  
$75  \le n \le 100$   &  1800   &  75 \% & 73 \% &  77 \%  & 76 \%  \\ 
$100 \le n \le 150$   &  1600   &  80 \% & 75 \% &  78 \%  & 77 \%  \\ 
$150 \le n \le 200$   &   700   &  81 \% & 79 \% &  80 \%  & 81 \%  \\ 
\hline \hline 
\end{tabular}
\caption[]{Identification power $\epsilon$ for gamma and protons
showers hitting in the centre of the detector (third and fourth colummns), 
and on the edge of the detector (last two colummns), for different pad-number 
intervals.}  
\label{tab:results}  
\end{table}

In order to study the effects of the core position, we generated a set of 
events with the core on the edge of the detector.
Also in this case the neural network rejection power is slightly affected,
and this fact can be understood by underlining that 
even for events hitting the apparatus at one edge, a fraction of the hit can 
be detected at great distances from the center and this, at
least for high populated events, can balance the loss of information 
due to the part of the shower hitting outside the detector.
The effect of the core reconstruction leads now to a reduction of the
discrimination powers $\epsilon$ of the order of
few percent.
\par
We underline, however, that the discrimination power is generally 
of the order of $ 75 \ \%$ or more, a value high enough to obtain a 
substancial enhancement in the sensitivity of the detector. If we 
consider the number of data detected in a given direction, the sensitivity of 
the apparatus results from
the ratio between the gamma flux (unknown) and the 
fluctuation of the hadron showers background (a well known value). In a 
simplified form, this ratio is given by:

\begin{equation}
\frac{n(\gamma)}{\sqrt{n(h)}} 
\ = \
\frac
{
\phi(\gamma) A_{\gamma} T_{obs} d\Omega
}
{
\sqrt{\phi(h) A_{h} T_{obs} d\Omega}
} 
\ \simeq 
\  \frac
{\phi(\gamma)}{\sqrt{\phi(h)}} 
\ \sqrt{A T_{obs} d\Omega}
\label{eq:ratiosim}
\end{equation}
With the aid of the neural network rejection, the number of protons is 
suppressed by a factor $(1-\epsilon_{h})$ (neglecting, of course,
the contribution from the gamma showers), while the gamma signal is reduced
by a factor $\epsilon_{\gamma}$.
\begin{wrapfigure}{r}{8.cm}
\centerline{\epsfig{figure=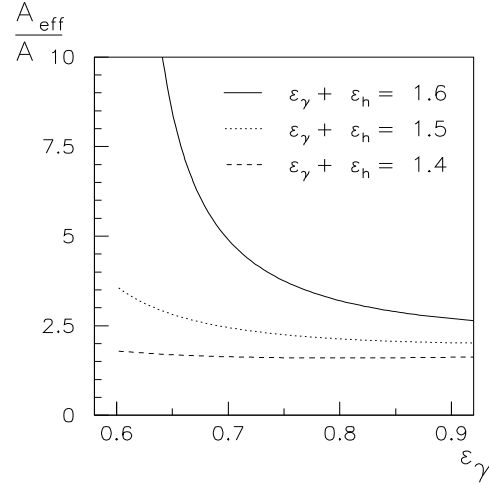,width=6.7 cm,height=6.5cm}}
\caption[]{The enhancement in the gamma ray sensitivity, as
a function of $\epsilon_{\gamma}$, for three different values of
the global identification power of the network: $80 \%$ (full line), 
$75 \%$ (dotted line) and $70 \%$ (dashed line).} 
\vspace{0.4cm}
\label{fig:globres}
\end{wrapfigure}
The previous equation becomes:
\begin{equation}
\frac{n(\gamma)}{\sqrt{n(h)}}  
\ \simeq
\  \frac{\phi(\gamma)}{\sqrt{\phi(h)}} 
\  \frac{\epsilon_{\gamma}}{\sqrt{(1-\epsilon_{h})}} 
\ \sqrt{A T_{obs} d\Omega} 
\label{eq:ratiorid}
\end{equation}
\par

This corresponds to an effective area given by:
\begin{equation}
\frac{A_{eff}}{A} \ \simeq \ \frac{\epsilon(\gamma)^{2}}{1-\epsilon_{h}}
\label{eq:reaeff}
\end{equation}
Taking, from
table \ref{tab:results}, $\epsilon_{\gamma} \simeq \epsilon_{h} \simeq 0.8$,
we obtain:
\begin{equation}
\frac{A_{eff}}{A} \ \simeq \ 3
\label{eq:reanum}
\end{equation}
This means that a rejection factor of the order of $80 \ \% $ furnishes  an 
increase in the sensitivity corresponding to an apparatus three times bigger, 
while a rejection factor of the order of $70 \ \% $, which represents our worst
result, corresponds to an apparatus more than two times bigger.  
We underline that the rejection power of the neural network can be tuned in 
order to fit the particular problem under analysis. In the
field of gamma astronomy, the relevant quantity is that of eq. \ref{eq:ratiorid}. In fig.
\ref{fig:globres} we show the behavior of this quantity, as a function of 
$\epsilon_{\gamma}$ for different values of 
$\epsilon_{\gamma} + \epsilon_{h}$ 
which represents an estimation of the global identification power of 
the neural network.
\par
Therefore, the use of this neural network improves the sensitivity of the 
apparatus, lowering the total observation time required to detect a gamma 
ray source above the cosmic ray background.

%
%
%
%
%
\vspace{1ex}
\begin{center}
{\Large\bf References}
\end{center}
%
Abbrescia, M. et al., 1996, Astroparticle physics with
Argo (unpublished). This document can be downloaded at
the URL: http://www1.na.infn.it/wsubnucl/cosm/argo/argo.html\\
D'Ettorre, B. et al., 1999, Proc. 26th ICRC (Salt Lake City, 1999) \\
Fesefeldt, H., 1985, PITHA Report  85-02 (Reading: RWTH Aachen) \\
Heck, D. et al., 1998, FZKA Report 6019
(Reading: Forschungszentrum Karlsruhe GmbH) \\
Knapp, J., Heck, D. \& Schatz, G., 1996, FZKA Report 5828
(Reading: Forschungszentrum Karlsruhe GmbH) \\
Mari, S.M. et al. 1999, Proc. 26th ICRC (Salt Lake City, 1999) \\
Nelson, W.R., Hirayama, H. \& Rogers, D.W., 1985, SLAC Report  265 
(Reading: Stanford National Acceleration Center) \\
Werner, K., 1993, Phys. Rep. 232, 87
\end{document}